\documentclass[aps,prd,reprint,preprintnumbers,superscriptaddress,nofootinbib]{revtex4-1}
\usepackage[all]{xy}
\usepackage{amsmath,amsthm,amssymb}
\usepackage[dvipsnames, usenames]{xcolor}
\usepackage{comment}
\usepackage{array}
\usepackage{bm}
\usepackage[normalem]{ulem} 
\allowdisplaybreaks[1]

\usepackage{hyperref}
\usepackage{tcolorbox}
\usepackage{mathrsfs}
\usepackage{float}
\usepackage{orcidlink}
\usepackage{tikz}
\usepackage{xspace}

\newcommand{\Eq}[1]{Eq.~\eqref{eq:#1}}

\newcommand{\spec}{\textsc{SpEC}\xspace}

\begin{document}

\title{Black-hole scattering with numerical relativity: \\ Self-force extraction and post-Minkowskian validation}

\def\AEI{Max Planck Institute for Gravitational Physics (Albert Einstein Institute), D-14476 Potsdam, Germany}
\def\Cornell{Cornell Center for Astrophysics and Planetary Science, Cornell University, Ithaca, New York 14853, USA}

\author{Oliver Long \orcidlink{0000-0002-3897-9272}}
\email{oliver.long@aei.mpg.de}
\affiliation{\AEI}
\author{Harald P. Pfeiffer \orcidlink{0000-0001-9288-519X}}
\affiliation{\AEI}
\author{Lawrence E. Kidder \orcidlink{0000-0001-5392-7342}}
\affiliation{\Cornell}
\author{Mark A. Scheel \orcidlink{0000-0001-6656-9134}}
\affiliation{TAPIR 350-17, California Institute of Technology, 1200 E California Boulevard, Pasadena, California 91125, USA}

\date{\today}
\begin{abstract}
The asymptotic nature of unbound binary-black-hole encounters provides a clean method for comparing different approaches for modeling the two-body problem in general relativity. In this work, we use numerical relativity simulations of black-hole scattering, generated using the Spectral Einstein Code, to explore the self-force and post-Minkowskian expansions of the scattering angle. First, we use a set of unequal-mass simulations to extract the self-force contributions to the scattering angle. Our main result is that using information up to second-order in the symmetric mass ratio (2SF) reproduces numerical relativity within the error bars across the full range of mass-ratios, including equal mass. Next, we compare our numerical relativity results to state-of-the-art post-Minkowskian predictions at larger impact parameters than previously explored. We find good agreement in the weak-field regime and discuss the relative importance of higher order terms.
\end{abstract}

\maketitle

\section{Introduction}

The study of black hole scattering has recently garnered significant interest as a novel avenue for probing the dynamics of strong-field gravity and enhancing our understanding of binary-black-hole (BBH) systems. The asymptotic nature of the scattering problem allows for the definition of gauge-invariant initial conditions, enabling unambiguous comparisons of key observables, such as the scattering angle, which can be directly compared across different theoretical frameworks and numerical simulations.

The most natural description of small deflection scattering is through the post-Minkowskian (PM) expansion, which treats the gravitational interaction perturbatively in powers of Newton's constant $G$. The vast majority of recent theoretical efforts in this area have focused on calculating high-order PM contributions to scattering observables using a variety of techniques borrowed from high-energy physics, including scattering amplitudes \cite{Bern:2021dqo,Bern:2021yeh,Damgaard:2023ttc} and worldline field theory \cite{Kalin:2019rwq,Dlapa:2021vgp,Dlapa:2022lmu,Dlapa:2023hsl,Jakobsen:2023ndj,Jakobsen:2023pvx,Driesse:2024feo,Driesse:2024xad}. These calculations have provided valuable insights into the dynamics of BBH encounters and have been used to construct bound-orbit waveform models~\cite{Buonanno:2024byg,Damour:2025uka}. However, these calculations break down when approaching the strong field regime, where the curvature of spacetime becomes significant, and non-perturbative effects dominate the dynamics. In order to remain faithful at small separations, one can apply resummation techniques such as effective-one-body (EOB)~\cite{Damour:2022ybd,Buonanno:2024vkx,Buonanno:2024byg,Damour:2025uka,Long:2025nmj,Clark:2025kvu} or incorporate information about the scatter-capture separatrix~\cite{Damour:2022ybd,Long:2024ltn}.

A different modeling approach is to expand the spacetime perturbatively in the mass ratio of the two black holes, which is known as the self-force (SF) expansion. This method is valid for arbitrary separations provided that the mass ratio is sufficiently small. While calculations of bound-orbit SF have reached exquisite accuracy~\cite{Wardell:2021fyy,Warburton:2021kwk,Mathews:2025txc}, hyperbolic SF calculations have been restricted to the case of a scalar charge scattering off a black hole~\cite{Barack:2022pde,Whittall:2023xjp,Whittall:2025dqn}. This toy model has been used to show the complementarity between the SF and PM methods including the extraction of higher-order PM coefficients from numerical SF calculations~\cite{Barack:2023oqp} and the resummation of PM using SF information to produce a semi-analytic model accurate across all separations~\cite{Long:2024ltn}.

The PM and SF frameworks are perturbative by construction, and determining their accuracy requires comparison with fully non-perturbative results. Numerical relativity (NR) provides this benchmark by directly solving discretized versions of the Einstein field equations  on high-performance computers. These simulations provide a detailed, fully non-linear description of phenomena and capture all of the physics of the system. While NR codes were originally designed to model the late inspiral and merger of compact binaries, they have recently been adapted to study scattering encounters~\cite{Damour:2014afa,Damour:2022ybd,Hopper:2022rwo,Rettegno:2023ghr,Albanesi:2024xus,Swain:2024ngs,Fontbute:2025vdv,Pretorius:2007jn,Sperhake:2008ga,Witek:2010xi,
  Sperhake:2010uv,Sperhake:2012me,Sperhake:2015siy,Kankani:2024may,Nelson:2019czq,Jaraba:2021ces,Rodriguez-Monteverde:2024tnt,Bae:2023sww,Fontbute:2024amb,Long:2025nmj,Kogan:2025vml}.

In this paper, we push the NR calculations of BBH scattering further into the perturbative regimes and harness the full nonlinearity of the simulation to probe the PM and SF expansions. We find that we can extract up to 2SF contributions from a set of unequal-mass NR simulations and that these terms reproduce the NR results across the full range of mass ratios, including the equal-mass case. We also perform a new set of weak-field NR simulations at varying impact parameters to validate state-of-the-art PM predictions and gain insights into higher-order contributions. The structure of the paper is as follows. In Sec.~\ref{sec:NRSimulations} we summarize the NR code employed here, the Spectral Einstein Code (\spec{}) and our procedure for extracting the scattering angle from the NR simulations. In Sec.~\ref{sec:SFExtraction} we analyze a set of simulations of varying mass ratio at fixed (rescaled) energy and angular momentum to extract high-order SF contributions. In Sec.~\ref{sec:PMExtraction} we compare a new set of weak-field NR simulations of varying impact parameter at fixed energy and mass ratio to PM calculations, finding good agreement. We conclude in Sec.~\ref{sec:Conclusions} with a summary of our findings and provide an outlook.

\subsection*{Notation}

In this work, we use natural geometrized units with $G=c=1$. Our setup consists of two non-spinning BHs with masses $m_1$ and $m_2$ with $m_1 \geq m_2$.
We denote the total mass, mass ratio, and symmetric mass ratio by
\begin{align}
M = m_1 + m_2, \qquad
q = \frac{m_1}{m_2}, \qquad
\nu = \frac{m_1 m_2}{M^2}.
\end{align}
We use rescaled versions of the Arnowitt-Deser-Misner (ADM) energy and angular momentum, $E_{\rm ADM}$ and $J_{\rm ADM}$, defined by
\begin{align}
\Gamma = \: \frac{E_{\rm ADM}}{M} = \sqrt{1+2\nu(\gamma-1)}, \qquad
\ell = \: \frac{J_{\rm ADM}}{m_1 m_2},
\end{align}
where $\gamma$ is the relative Lorentz factor,
\begin{equation}\label{eq:gammaDef}
\gamma = \frac{\Gamma^2M^2 -m_1^2 - m_2^2}{2m_1m_2}.
\end{equation}
We define the impact parameter as
\begin{equation}\label{eq:bDef}
b = \frac{\ell \: \Gamma}{\sqrt{\gamma^2-1}} M. 
\end{equation}

\section{Numerical relativity simulations}
\label{sec:NRSimulations}

This work uses the Spectral Einstein Code
(\spec{})~\cite{SpECwebsite}, a multi-domain spectral code for
solving the general relativistic initial value and evolution problem for BBH systems.
The numerical techniques are summarized in
Refs.~\cite{Mroue:2013xna,Boyle:2019kee,Scheel:2025jct}.  In particular, \spec{}
evolves a first-order representation of the generalized harmonic
evolution system~\cite{Lindblom:2005qh} using a multi-domain spectral
method~\cite{Kidder:1999fv, Scheel:2008rj, Szilagyi:2009qz,
  Hemberger:2012jz}.  The outer boundary uses
  constraint-preserving  
boundary conditions~\cite{Lindblom:2005qh, Rinne:2006vv, Rinne:2007ui}
with (approximately) no incoming gravitational radiation.
Black hole (BH) excision is used inside the apparent
horizons~\cite{Scheel:2008rj, Szilagyi:2009qz, Hemberger:2012jz,
  Ossokine:2013zga}.  The elliptic solver of
\spec{}~\cite{Pfeiffer:2002wt,Ossokine:2015yla} utilizes the
extended conformal-thin sandwich (XCTS) approach
\cite{York:1998hy,Pfeiffer:2002iy,Cook:2004kt}. 

\spec{} was recently adapted to perform accurate simulations of hyperbolic encounters. These changes included modifications to the initial data routines, to account for larger separations and velocities, as well as the adaptive mesh refinement procedure during the evolution. Details of these modifications can be found in Refs.~\cite{Long:2025nmj,Mendes:2025gov}.

\begin{figure}
\centering
\includegraphics[width=\linewidth,trim=7 7 7 6,clip=true]{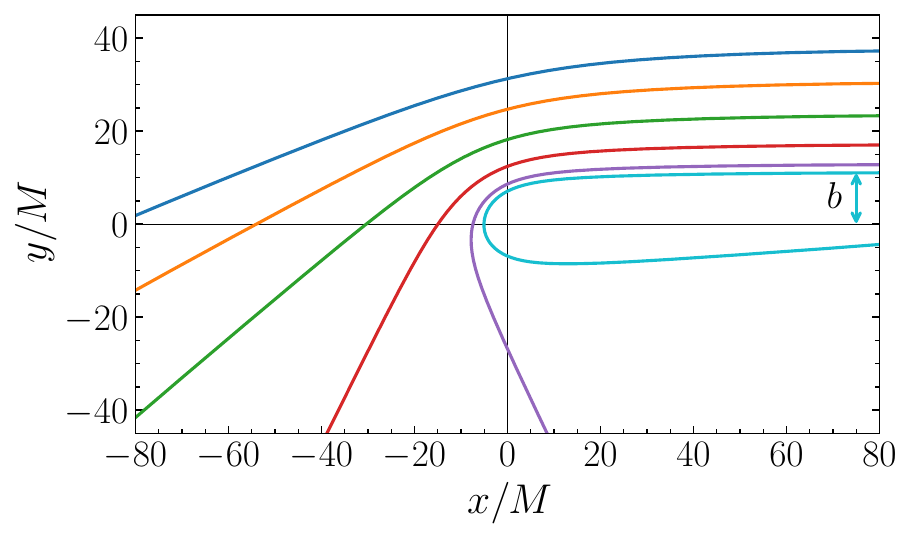}
\caption{ 
Trajectories of equal-mass black-hole scattering scenarios with $\Gamma=1.02264$ and impact parameters $b=[37.4, 30.4, 23.4, 17.1, 12.8, 11.1]M$. Shown are the Cartesian components of the separation vector between the centers of the BHs, rotated such that the incoming path is aligned with the $x$-axis. The $b=17.1M$ (red) trajectory corresponds to the largest impact parameter previously studied at this energy~\cite{Damour:2014afa,Rettegno:2023ghr,Swain:2024ngs,Long:2025nmj}.
}\label{Fig:Traj}
\end{figure}

In this study we analyze the observables from 23 simulations previously presented in Ref.~\cite{Long:2025nmj}, combined with 9 new equal-mass simulations at fixed $\Gamma=1.02264$ deeper in the weak field (impact parameters $\ell \geq 8$ corresponding to $b \geq 18.7M$). A subset of the equal-mass simulations are shown in Figure~\ref{Fig:Traj}. The rest of this section is dedicated to summarizing the details of the simulations performed and our procedure for extracting the scattering angle from the data.

\subsection{Simulations}

The setup of all simulations used here is exactly as detailed in Sec.~II of Ref.~\cite{Long:2025nmj}, to summarize:  

Each simulation is specified by the parameters of the BHs (masses and spins) combined with parameters that determine the trajectories of the binary, notably the ADM energy and angular momentum, $E_{\rm ADM}$ and $J_{\rm ADM}$, as well as the initial separation $D_0=250M$. In order to evolve this system, we solve for initial data on the first slice of the evolution. The \spec{} initial data routine \cite{Pfeiffer:2002wt,Ossokine:2015yla} performs a root-finding procedure over additional parameters to achieve the desired system. This includes those that control the initial position and velocity of the BHs as well as the conformal metric, trace of the extrinsic curvature and their time-derivatives and the boundary conditions at the BH horizons and spatial infinity. As discussed in Ref.~\cite{Long:2025nmj}, we can achieve high accuracy in the desired values of the masses, spins, ADM energy and angular momentum, etc., with relative errors $\lesssim 10^{-5}$, ensuring that the results here are not affected by uncertainty in the initial parameters.

The evolution starts with the BHs at Cartesian coordinates $\vec c_A \approx (D_0/(1+q), 0, 0)$ and $\vec c_B \approx (-D_0 \, q/(1+q), 0, 0)$, corresponding to the center-of-mass frame (GR corrections cause small deviations relative to these Newtonian values, see Ref.~\cite{Ossokine:2015yla}). Because we restrict to non-spinning BHs, the motion of the BHs is confined to the $xy$-plane. The simulation is terminated when the BHs reach a separation of $1.4 D_0$ on the outgoing leg, beyond which \spec's grid-decomposition becomes inaccurate.

\subsection{Scattering angle extraction}

The scattering angle $\theta$ is a gauge-invariant quantity that can be used to compare different simulations and analytic predictions. In order to calculate this quantity we must track the motion of each BH throughout the simulation and extrapolate their trajectories to infinite separation. Here we review the extrapolation procedure and refer the reader to Sec.~II of Ref.~\cite{Long:2025nmj} for a detailed explanation.

Due to the trajectories being confined to the $xy$-plane, the asymptotic motion of the BHs are specified by the ingoing and outgoing azimuthal angles $\varphi^{\infty,{\rm in}}$ and $\varphi^{\infty,{\rm out}}$ respectively. The scattering angle $\theta_i$ of the $i$th BH is defined in terms of the asymptotic values of the azimuthal angles by
\begin{equation}
\label{eq:def_scattering_angle}
\theta_i = \varphi^{\infty,{\rm out}}_i - \varphi^{\infty,{\rm in}}_i - \pi.
\end{equation}

For $q=1$, symmetry under exchange of the two bodies implies $\theta_1=\theta_2$; for unequal masses, however, asymmetric momentum emission (a BH kick) can result in $\theta_1\neq \theta_2$.
The motion of each BH is tracked as the Cartesian coordinate of the center of the excised region in the initial center-of-mass frame of the system. This is converted into polar coordinates $(r,\varphi)$ on each leg of the orbit. In order to extrapolate the azimuthal angle to infinite separation we fit $\varphi$ to a Keplerian ansatz
\begin{equation} \label{eq:KeplerFit}
\frac{1}{D} = A \cos \varphi + B \sin\varphi + C,
\end{equation}
where $A$, $B$, and $C$ are all free fitting parameters.
Equation~(\ref{eq:KeplerFit}) is equivalent to the standard form of a Keplerian orbit,
\begin{equation}\label{eq:Kepler}
D = \frac{p}{1+e\cos(\varphi-\varphi_0)},
\end{equation}
with the identifications
\begin{align}
p =&\: \frac{1}{C},\\
e =&\: \frac{\sqrt{A^2+B^2}}{C},\\
\varphi_0 =&\: \arctan (A/B).
\end{align}
In addition, we also use a more agnostic fit of the form
\begin{equation} \label{eq:PolyFit}
\varphi  = \varphi^\infty + \frac{\varphi^{(1)}}{D} + \frac{\varphi^{(2)}}{D^2} + \frac{\varphi^{(3)}}{D^3},
\end{equation}
where $\varphi^\infty$, $\varphi^{(i)}$ are free fitting parameters.
Here the asymptotic azimuthal angle is the constant term, $\varphi^\infty$, whereas in the Keplerian fit it can be recovered with
\begin{equation}\label{eq:KeplerAngle}
\varphi^\infty = \varphi_0 \pm \arccos(-1/e),
\end{equation}
where the $+$ and $-$ signs correspond to the ingoing and outgoing legs respectively.

The fits for both models are performed over the portion of the trajectory, for which $D\in[d_{\rm in}, 200M]$ with $d_{\rm in}\in[50M, 80M]$ for the ingoing legs and $D\in[d_{\rm out}, 350M]$ with $d_{\rm out}\in[80M, 230M]$ for the outgoing legs. The different treatment of the ingoing leg allows unphysical junk-radiation arising from the initial data to dissipate. The final value of $\varphi^\infty$ is taken as the average over all fits, with the uncertainty estimated from the spread of values. The final scattering angle is then calculated using \Eq{def_scattering_angle}, with the uncertainty estimated by adding the uncertainties in quadrature.

In this paper we only consider the average scattering angle
\begin{equation}
\theta = \frac{\theta_1 + \theta_2}{2}.
\end{equation}
For the equal mass case considered in Sec.~\ref{sec:PMExtraction}, this is equivalent to the individual scattering angles, whereas for the unequal mass case in Sec.~\ref{sec:SFExtraction}, the broken symmetry of the system and the associated asymmetric gravitational wave emission means that $\theta_1 \neq \theta_2$. However, as shown in Ref.~\cite{Long:2025nmj}, the difference in the angles of each BH is small compared to the overall scattering angle ($\lesssim 0.1\%$) and thus we neglect the difference here.

\section{Self-force regime}
\label{sec:SFExtraction}

The SF expansion\footnote{Formally the SF is an expansion in the small mass ratio $\varepsilon = m_2/m_1$, however, comparisons between SF and NR have shown better agreement at comparable masses when resumming the SF expansion in terms of the symmetric mass ratio $\nu = \varepsilon/(1+\varepsilon)^2$.} of the scattering angle is given by
\begin{equation}
\theta = \vartheta_{\rm 0SF} + \nu \vartheta_{\rm 1SF} + \nu^2 \vartheta_{\rm 2SF} + \nu^3 \vartheta_{\rm 3SF} + \ldots\,.
\label{eq:SFAngle}
\end{equation}
The leading-order term $\vartheta_{\rm 0SF}$ represents the scattering angle of a geodesic in a Schwarzschild spacetime and can be calculated using elliptic integrals~\cite{Barack:2022pde}. The higher-order terms, $\vartheta_{n\text{SF}}$ for $n\geq 1$, represent corrections due to the finite mass ratio of the binary system. These terms are significantly more challenging to compute and thus have only been calculated with additional approximations, such as the latest PM results which contain information at $G^5$ up to linear order in the mass ratio, i.e. 5PM(1SF)~\cite{Driesse:2024xad,Driesse:2024feo}.

Here we explore an alternative method of calculating the self-force coefficients by using our NR simulations to extract higher-order SF contributions to the scattering angle. The simulations used here cover mass ratios $q\in[1,10]$ at fixed $\gamma=1.0200$ and $\ell=4.8000$\footnote{By fixing $\gamma$ and $\ell$ we ensure that the configurations smoothly approach the geodesic limit as $\nu \rightarrow 0$.}.

\begin{figure*}
\centering
\includegraphics[width=\linewidth,trim=7 11 8 11,clip=true]{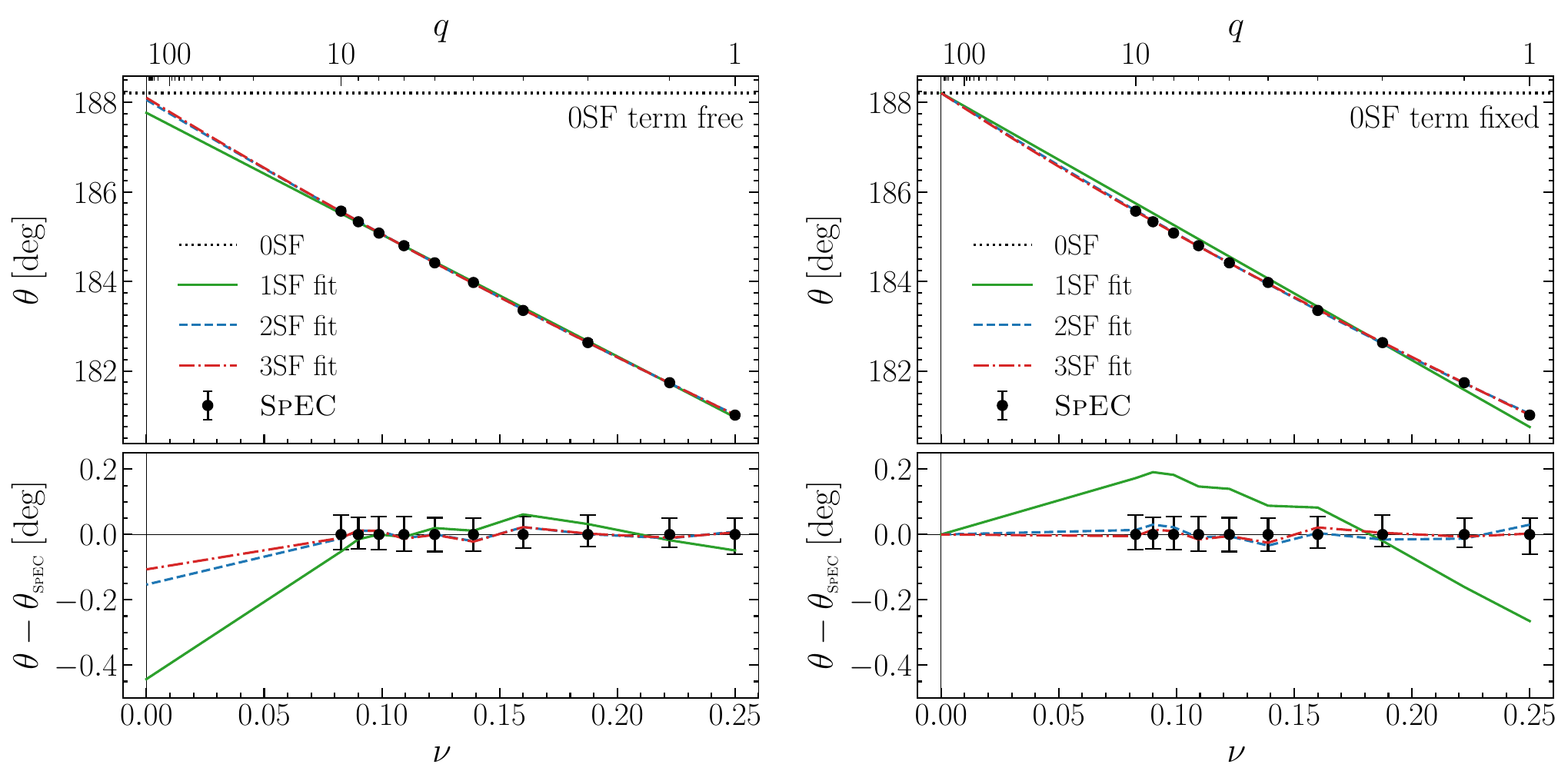}
\caption{ 
Scattering angle $\theta$ as a function of symmetric mass ratio $\nu$ at fixed $\gamma=1.0200$ and $\ell=4.8000$. The points represent the \spec{} data and the lines represent the best fits to the data using Eq.\ (\ref{eq:SFAngle}). The $n$SF fit contains all terms up to $n$th order in $\nu$ with the extracted values shown in Table~\ref{Table:SFAngle}. The known 0SF term (dotted black line) and the zero mass limit (thin vertical line) are shown for reference. {\em Left:} Fits where all SF coefficients are free parameters (Fits 1--3 of Table~\ref{Table:SFAngle}). {\em Right:} Fits where the 0SF term is fixed to the known geodesic value $\vartheta_{\rm 0SF}=188.211^\circ$ (Fits 1'--\:3' of Table~\ref{Table:SFAngle}).
}\label{Fig:SFFits}
\end{figure*}

We perform the extraction by fitting the numerical data for $\theta$ to the SF ansatz (\ref{eq:SFAngle}) and varying the number of fitting parameters to ensure the extracted results are consistent (left panel of Fig.~\ref{Fig:SFFits}). We also perform fits fixing the known leading-order geodesic (0SF) coefficient to further ensure consistency between the fitted values (right panel of Fig.~\ref{Fig:SFFits}). One important observation from Figure \ref{Fig:SFFits} is that all of the fits describe the data well with the maximum deviation between the data and fits being $\sim 0.4^\circ$. However, it is apparent that the 1SF fits do not contain enough information to describe the data across the full range of mass ratios to within the NR errors. When $\vartheta_{0\rm SF}$ is fitted (left panel of Fig.~\ref{Fig:SFFits}), the 1SF fit
recovers $\vartheta_{0\rm SF}$ with a comparatively large error of $0.4^\circ$ compared to the known geodesic value at $\nu=0$. Similarly, when $\vartheta_{0\rm SF}$ is fixed to the known value (right panel of Fig.~\ref{Fig:SFFits}), the 1SF fit does not closely match the data. This indicates that higher-order SF terms are required to accurately describe the data across the full range of mass ratios, which is confirmed by the 2SF and 3SF fits shown in Fig.~\ref{Fig:SFFits}. Because the 2SF fits already remain within the NR error bars, the 3SF fit adds no information over the 2SF fit and thus we are unable to extract any information about the 3SF coefficient.

\begin{table}[tb]
\begin{center}
\begin{tabular}{c|l|l|l|l}
\hline
Fit $\#$ & \multicolumn{1}{c|}{$\vartheta_{\rm 0SF}$}            & \multicolumn{1}{c|}{$\vartheta_{\rm 1SF}$}           & \multicolumn{1}{c|}{$\vartheta_{\rm 2SF}$}       & \multicolumn{1}{c}{$\vartheta_{\rm 3SF}$}          \\ \hline
1 & $187.77(3)$     &       $-27.2(2)$          &      \multicolumn{1}{c|}{--}      &          \multicolumn{1}{c}{--}       \\ 
2 & $188.06(4)$     &       $-31.2(7)$          &      $12(2)$       &          \multicolumn{1}{c}{--}       \\ 
3 & $188.1(2)$     &       $-32(4)$          &      $19(24)$       &          $-12(50)$       \\ \hline
1' & ${\bf 188.211}$     &       $-29.8(3)$          &      \multicolumn{1}{c|}{--}       &          \multicolumn{1}{c}{--}       \\ 
2' & ${\bf 188.211}$     &       $-33.3(2)$          &      $18.5(9)$       &          \multicolumn{1}{c}{--}       \\ 
3' & ${\bf 188.211}$     &       $-34.4(4)$          &      $32(5)$       &          $-40(14)$       \\ 
\hline
\end{tabular}
\caption{
\label{Table:SFAngle}
Values of the SF coefficients $\vartheta_{n\text{SF}}$ (in degrees) obtained by fitting the numerical data for $\theta$ to the SF model (\ref{eq:SFAngle}). In each row, numerical entries represent fitted terms, with parentheses representing the uncertainty in the last digit(s). \textbf{Bold} entries indicate terms that are fixed to the known 0SF value and dashes indicate terms not fitted.
}
\end{center}
\end{table}

\begin{figure}
\centering
\includegraphics[width=\linewidth,trim=7 11 8 12,clip=true]{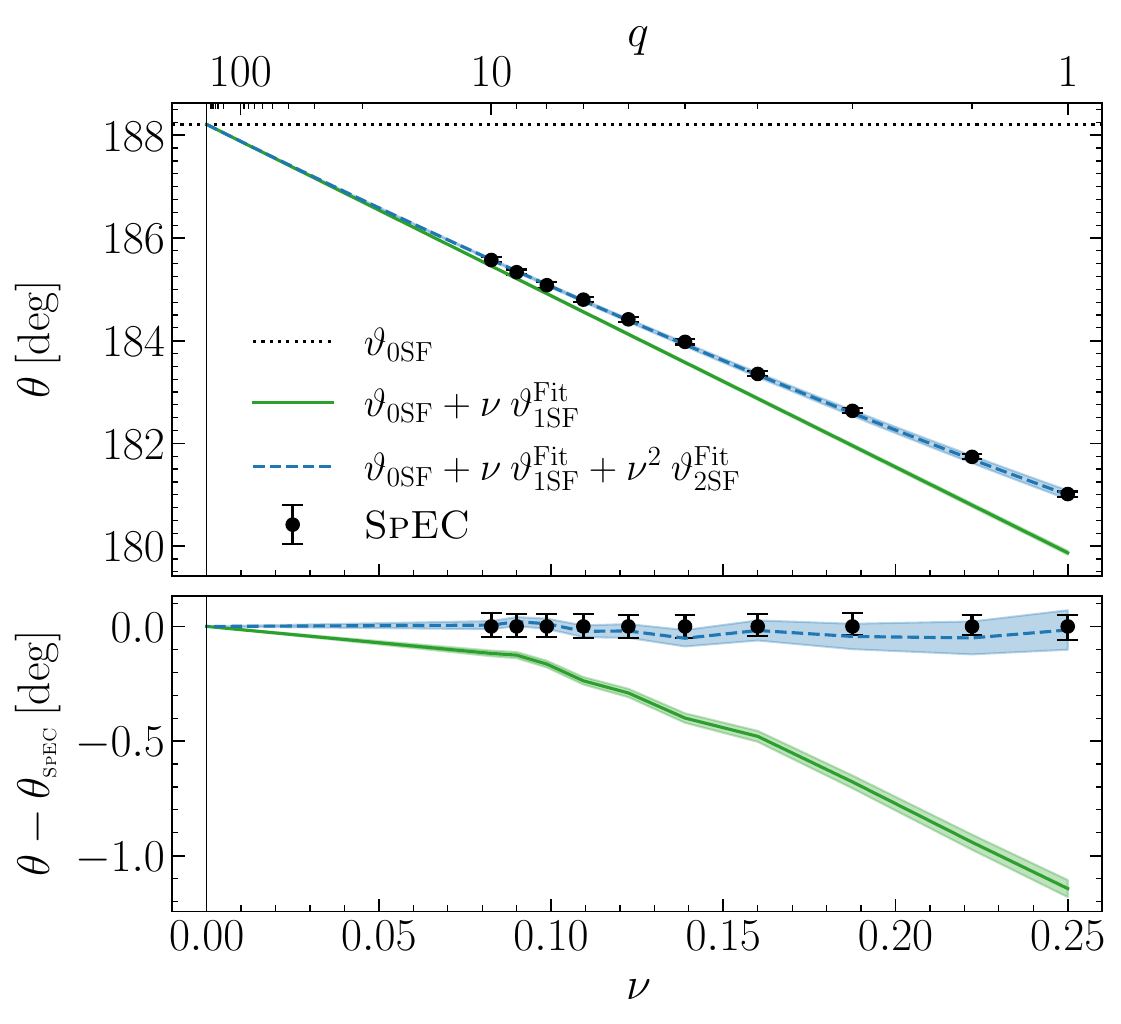}
\caption{ 
  Scattering angle $\theta$ as a function of symmetric mass ratio $\nu$ at fixed $\gamma=1.0200$ and $\ell=4.8000$. The points represent the \spec{} data and the lines correspond to the extracted SF expansion of Eq.\ (\ref{eq:SFAngle}) using the best-estimate coefficients in Eqs.~(\ref{eq:1SFFit})--(\ref{eq:2SFFit}) and the known 0SF term. The known 0SF term (dotted black line) and the zero mass limit (thin vertical line) are shown for reference. The lower panel shows the difference between the data and the best-estimate fits.}
\label{Fig:SFAvgFits}
\end{figure}

We can use the values of the fitted coefficients (shown in Table~\ref{Table:SFAngle}) to extract estimates for each SF coefficient. We calculate the final estimates for each SF coefficient by taking the average of the fitted values of each coefficient weighted by the square of the inverse of their uncertainties.
However, as we know that the 1SF term is insufficient to describe the data across the full range of mass ratios, we only consider the 2SF and 3SF fits for this extraction, i.e. Fits 2, 2', 3, and 3' from Table \ref{Table:SFAngle}.
Applying this procedure, we find the best estimates for the SF coefficients to be
\begin{align}
\vartheta_{\rm 1SF}^{\rm Fit} &\:= -33.36\pm 0.15, \label{eq:1SFFit} \\
\vartheta_{\rm 2SF}^{\rm Fit} &\:= \phantom{-}18.0\pm 0.8. \label{eq:2SFFit}
\end{align} 
We do not provide the value of $\vartheta_{\rm 3SF}^{\rm Fit}$ here because, as discussed earlier, both the 2SF and 3SF fits are entirely within the NR errors, so we are unable to extract meaningful information about the 3SF coefficient. Figure \ref{Fig:SFAvgFits} shows the numerical data and the SF fits using the known geodesic value and the best estimate values with the shading representing the uncertainty in the fit. As can be seen, the 1SF fit is a good representation of the data at low $\nu$ but the deviation grows to $\sim 1^\circ$ at equal masses. The 2SF fit is a significant improvement over the 1SF fit and remains within the error bars across the full range of mass ratios. This demonstrates that NR data can be used not only to validate perturbative results but also to determine new coefficients, providing a non-perturbative route to infer higher-order information.

\section{post-Minkowskian regime}
\label{sec:PMExtraction}

The PM expansion of the scattering angle is given by
\begin{equation}
\theta = \sum_{n=1}^\infty \vartheta_{n\text{PM}} \left(\frac{GM}{b}\right)^n,
\label{eq:PMAngle}
\end{equation}
where $\vartheta_{n\text{PM}}$ is the $n$th order PM contribution to the scattering angle and we have restored Newton's constant $G$ for clarity. The state-of-the-art results are at 5PM(1SF) which contain all information up to $G^4$ \cite{Bern:2021dqo,Dlapa:2022lmu} as well as the contributions up to order $\nu$ at $G^5$~\cite{Driesse:2024xad,Driesse:2024feo}. In addition, as the $\nu^0$ contributions are purely geodesic, we have these contributions for all orders in $G$.

\begin{figure}
\centering
\includegraphics[width=\linewidth,trim=6 7 6 7,clip=true]{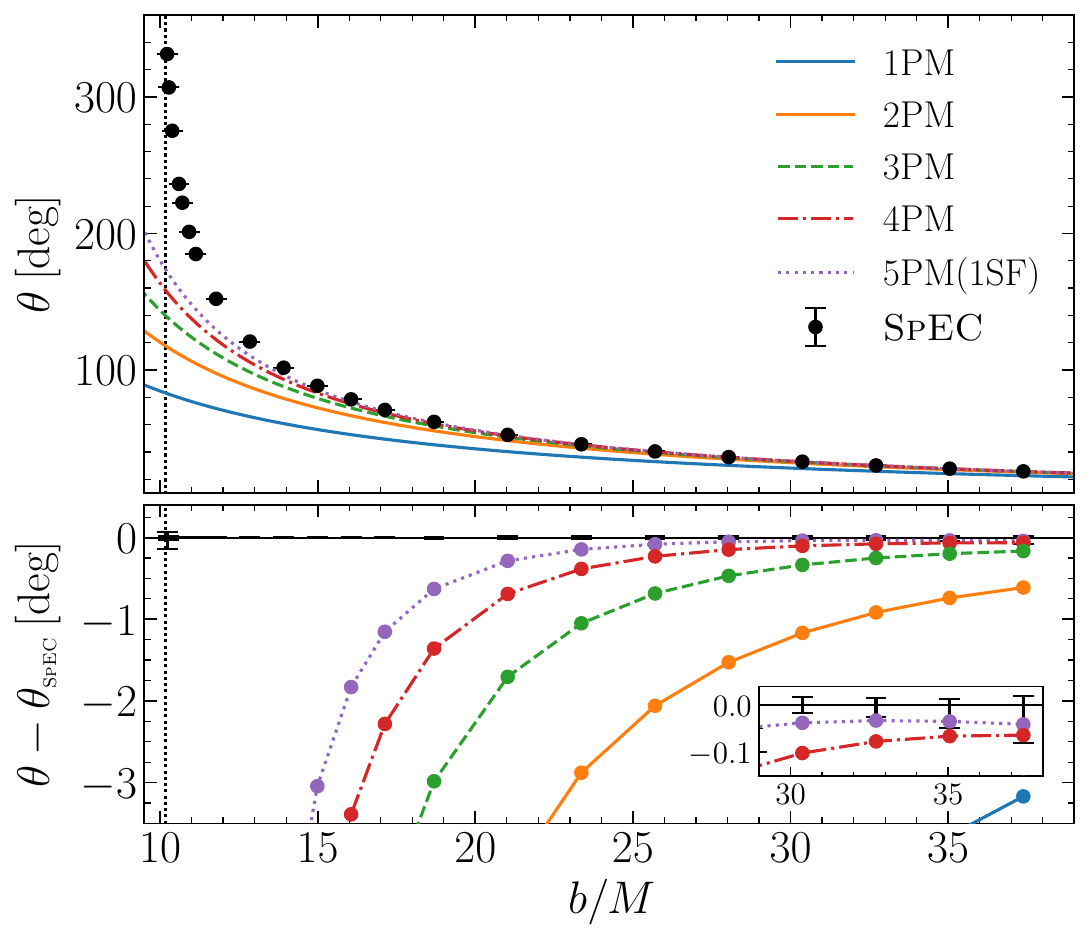}
\caption{ 
Scattering angle $\theta$ as a function of impact parameter $b$ at fixed $\Gamma=1.02264$ and $q=1$. The points represent the \spec{} data and the lines represent the analytic PM predictions. The vertical dotted lines show the first confirmed capture from \spec{}. The bottom panel shows the difference between the data and PM predictions with the inset showing a zoomed-in view of the weak-field region. The \spec errors are too small to be seen on the scale of the main plot, but can be resolved in the inset.
}\label{Fig:PMComp}
\end{figure}

Here we seek to utilize \spec's ability to simulate systems with large impact parameters to validate state-of-the-art PM calculations. The simulations used here cover impact parameters $b\in[10.2,37.4]M$ (including 9 new simulations with $b\geq 18.7M$) at fixed $\Gamma=1.02264$ and $q=1$. Figure~\ref{Fig:PMComp} shows the scattering angle $\theta$ as a function of impact parameter $b$ along with the PM predictions at various orders. As can be seen, the NR results agree well with the PM predictions at large $b$ (weak field) and deviate more significantly at small $b$ (strong field) where the PM expansion does not capture the divergent behavior of the scattering angle near the scatter-capture separatrix. The lower panel of Figure~\ref{Fig:PMComp} shows the difference between the NR results and the PM predictions. The inset of the lower panel shows that the state-of-the-art 5PM(1SF) prediction agrees with the NR results to within the error bars for $b\gtrsim 35M$ with the 4PM prediction agreeing within the error bars for $b\sim 37M$. This provides a strong validation of both the NR simulations and the PM calculations in the weak-field regime.

\begin{figure}
\centering
\includegraphics[width=\linewidth,trim=6 10 7 6,clip=true]{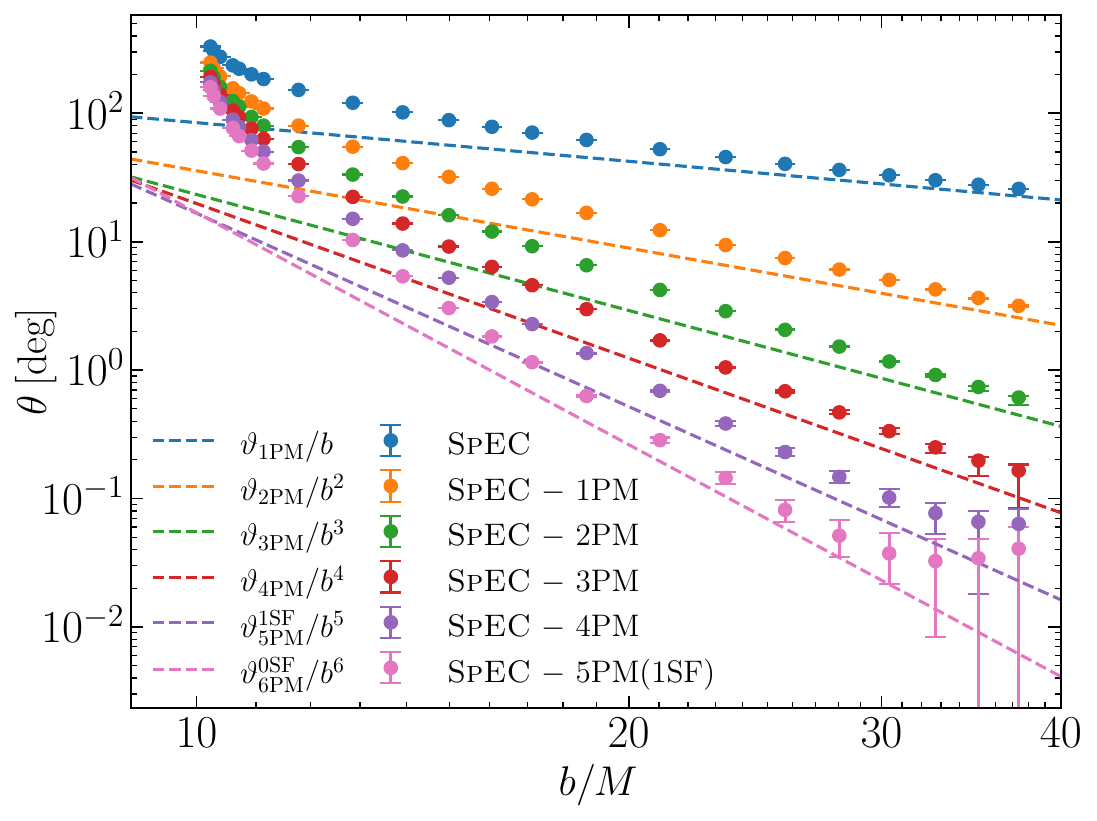}
\caption{ 
Scattering angle $\theta$ as a function of impact parameter $b$ at fixed $\Gamma=1.02264$ and $q=1$. The points represent the \spec{} data with various PM orders subtracted off, as indicated in the legend, and the dashed lines represent the analytic PM predictions at the corresponding order. For example, the orange points represent the \spec{} data with the 1PM prediction subtracted off and the orange dashed line represents the 2PM prediction. $\vartheta_{5\text{PM}}^{1\text{SF}}$ represents the coefficient of $G^5$ containing terms up to order $\nu$.
}\label{Fig:PMTrends}
\end{figure}

Another informative way to visualize the agreement between NR and PM is to plot the NR results with successive PM orders subtracted off, as shown in Figure~\ref{Fig:PMTrends}. If there is agreement, each set of points should tend towards the corresponding PM prediction with increasing $b$. As can be seen, the NR results agree well with the PM predictions at each order, providing a strong validation between the NR and PM calculations. One interesting observation is that when subtracting the 5PM(1SF) term, the points appear to follow a trend of $\sim 1/b^6$ at large $b$. This suggests that the next order term, 6PM, is larger than the missing 2SF contribution at 5PM. In fact, when comparing these results to the known 6PM $\nu^0$ contribution, we find that this is indeed the case and that the 6PM(0SF) term is actually within the error bars of the NR data at large $b$. While we expect that the $\nu^0$ contribution dominates the 6PM term, as has been observed at lower PM orders (see e.g.\ Fig.~7 of Ref.~\cite{Long:2025nmj}), this does not mean that the whole 6PM term is within the error bars as the unknown higher-order in $\nu$ contributions could still be significant.

With the NR data agreeing well with the PM predictions, it is natural to attempt to extract higher-order PM coefficients from the data. When we attempted this, the fits proved unstable as different combinations of fitted parameters led to large variations in the results. This likely occurs because we are not sufficiently deep into the weak-field regime for higher-order PM terms to be small compared with lower-order terms. Thus, we are unable to extract reliable higher-order PM coefficients from the current dataset.

To perform this extraction, we would require higher-accuracy simulations to ensure that the small differences between the NR results and PM predictions are not dominated by NR error, as well as simulations at larger impact parameters. However, as seen in Figs.~\ref{Fig:PMComp} and \ref{Fig:PMTrends}, increasing the impact parameter increases the error in scattering angle extraction. To reduce this error, we would need to scale the initial separation $D_0$ with the impact parameter which we leave to future work.

\section{Conclusions}
\label{sec:Conclusions}

In this work, we show how highly accurate NR simulations have become a new tool to explore perturbative methods and to predict higher order perturbative terms. First, we analyze an existing dataset which varies mass ratio in order to extract SF contributions to the scattering angle. We show that it is possible to extract the 1SF and 2SF contributions with high accuracy. The main result from this paper is that a comparison between the best-estimate value of the fits to the NR data shows that the 2SF fit is within the error bars across the full range of mass ratios considered, indicating that the 2SF contribution is sufficient to give the scattering angle to subpercent accuracy even at equal masses. This remarkable result mirrors the agreement between first post-adiabatic (1PA) SF results and NR in the quasicircular orbit case \cite{Wardell:2021fyy,Warburton:2021kwk,Mathews:2025txc}, demonstrating that SF methods have the potential to accurately model BBH systems even at comparable masses.

We also present a comparison between state-of-the-art PM predictions and a set of equal-mass NR simulations, including new simulations at larger impact parameters than previously explored. We find excellent agreement between the NR results and PM predictions in the weak-field regime, providing a strong validation of both approaches. The key finding from this comparison is that the 6PM(0SF) term appears to dominate over the unknown 5PM(2SF) term. Attempts were made to extract higher-order PM coefficients from the data, however, the fits were found to be unstable and thus we are unable to provide reliable estimates of these coefficients.

We can also compare how the SF and PM expansions perform beyond their respective regimes of validity. As shown in Figs.~\ref{Fig:SFAvgFits} and \ref{Fig:PMComp}, the SF expansion remains significantly more accurate than the PM one when moving away from the weak-field and small–mass-ratio limits. This can be attributed to the SF expansion encoding more physical information: for instance, terms up to 2SF include all contributions through 6PM order as well as all terms up to $\nu^2$ at higher orders in $G$~\cite{Damour:2019lcq}. Thus, the SF expansion is able to capture more of the dynamics of the full non-linear system, leading to better agreement with NR results.

This study presents the first comparisons between NR scattering data and perturbative expansions when approaching their regimes of validity. This opens up new avenues for future research, including exploring other observables, such as the radiated energy and angular momentum, and extending to other regions of parameter space such as spinning black holes. Additionally, further work is needed to extract higher-order PM coefficients from NR data, which may require simulations at even larger impact parameters to ensure convergence of the PM expansion. Overall, this work highlights the power of using NR simulations as a tool to both validate and inform perturbative methods in the study of BBH dynamics.

\section*{Acknowledgements}

The authors would like to thank O.~Markwell and P.J.~Nee for useful discussions. Computations were performed on the HPC system Urania at the Max Planck Computing and Data Facility. This work makes use of the Black Hole Perturbation Toolkit~\cite{BHPToolkit}.
This material is based upon work supported by the National Science Foundation under Grants No.~PHY-2407742, No.~PHY-2207342, and No.~OAC-2209655, and by the Sherman Fairchild Foundation at Cornell.
This work was supported by the National Science Foundation under Grants No.~PHY-2309211, No.~PHY-2309231, and No.~OAC-2513339 at Caltech, and the Sherman Fairchild Foundation at Caltech.

\section*{Data Availability}

An ancillary file containing the scattering angle data used in this article is openly available at \cite{Data}.

\appendix

\section{Numerical values}

In this appendix, we provide the numerical values of the scattering angles from our \spec{} simulations used in Secs.~\ref{sec:SFExtraction} and \ref{sec:PMExtraction}. The ancillary file \cite{Data} contains machine-readable versions of these tables with the exact initial conditions and extracted scattering angle values for each simulation.

\begin{table}[H]
    \centering
    \renewcommand{\arraystretch}{1.2}
    \begin{tabular}{c|c}
    \hline
$q$ & $\theta_{\spec{}}$ \\ \hline
$1.0$ & $181.01_{-0.06}^{+0.05}$ \\
$2.0$ & $181.74_{-0.04}^{+0.05}$ \\
$3.0$ & $182.63_{-0.04}^{+0.06}$ \\
$4.0$ & $183.35_{-0.04}^{+0.05}$ \\
$5.0$ & $183.98_{-0.05}^{+0.05}$ \\
$6.0$ & $184.42_{-0.05}^{+0.05}$ \\
$7.0$ & $184.80_{-0.05}^{+0.05}$ \\
$8.0$ & $185.08_{-0.05}^{+0.06}$ \\
$9.0$ & $185.33_{-0.04}^{+0.05}$ \\
$10.0$ & $185.57_{-0.05}^{+0.06}$ \\ \hline
    \end{tabular}
    \caption{Scattering angle (in degrees) for unequal mass, non-spinning BHs with $\gamma\!=\!1.0200$ and $\ell\!=\!4.8000$.}
\end{table}

\begin{table}
    \centering
    \renewcommand{\arraystretch}{1.2}
    \begin{tabular}{c|r|r|r|r|r|r}
    \hline
$b/M$ & \hspace{1.3em}$\theta_{\spec{}}$\hspace{1.3em} & \hspace{0.1em}$\theta_{1\text{PM}}$\hspace{0.1em} & \hspace{0.3em}$\theta_{2\text{PM}}$\hspace{0.3em} & \hspace{0.3em}$\theta_{3\text{PM}}$\hspace{0.3em} & \hspace{0.3em}$\theta_{4\text{PM}}$\hspace{0.3em} & \hspace{0.3em}$\theta^{1\text{SF}}_{5\text{PM}}$\hspace{0.3em} \\ \hline
$10.225$ & $331.395_{-0.145}^{+0.074}$ & $82.74$ & $116.92$ & $138.73$ & $156.88$ & $171.77$ \\
$10.279$ & $306.963_{-0.024}^{+0.017}$ & $82.31$ & $116.13$ & $137.61$ & $155.38$ & $169.89$ \\
$10.386$ & $275.229_{-0.008}^{+0.007}$ & $81.47$ & $114.59$ & $135.41$ & $152.46$ & $166.24$ \\
$10.600$ & $236.282_{-0.006}^{+0.006}$ & $79.82$ & $111.62$ & $131.21$ & $146.92$ & $159.36$ \\
$10.707$ & $222.541_{-0.006}^{+0.007}$ & $79.02$ & $110.19$ & $129.19$ & $144.28$ & $156.11$ \\
$10.921$ & $201.202_{-0.006}^{+0.007}$ & $77.47$ & $107.43$ & $125.34$ & $139.28$ & $150.00$ \\
$11.135$ & $184.967_{-0.006}^{+0.006}$ & $75.98$ & $104.80$ & $121.69$ & $134.59$ & $144.32$ \\
$11.778$ & $152.153_{-0.007}^{+0.007}$ & $71.84$ & $97.60$ & $111.87$ & $122.18$ & $129.53$ \\
$12.849$ & $120.848_{-0.008}^{+0.008}$ & $65.85$ & $87.49$ & $98.49$ & $105.77$ & $110.52$ \\
$13.919$ & $101.726_{-0.008}^{+0.008}$ & $60.79$ & $79.23$ & $87.88$ & $93.16$ & $96.35$ \\
$14.990$ & $88.443_{-0.009}^{+0.009}$ & $56.44$ & $72.35$ & $79.27$ & $83.20$ & $85.40$ \\
$16.061$ & $78.532_{-0.010}^{+0.010}$ & $52.68$ & $66.53$ & $72.16$ & $75.14$ & $76.70$ \\
$17.131$ & $70.787_{-0.010}^{+0.011}$ & $49.39$ & $61.56$ & $66.20$ & $68.51$ & $69.63$ \\
$18.692$ & $62.047_{-0.011}^{+0.012}$ & $45.27$ & $55.49$ & $59.06$ & $60.69$ & $61.42$ \\
$21.029$ & $52.529_{-0.013}^{+0.014}$ & $40.24$ & $48.32$ & $50.82$ & $51.84$ & $52.24$ \\
$23.365$ & $45.636_{-0.014}^{+0.015}$ & $36.21$ & $42.76$ & $44.59$ & $45.25$ & $45.49$ \\
$25.702$ & $40.388_{-0.016}^{+0.016}$ & $32.92$ & $38.33$ & $39.70$ & $40.16$ & $40.31$ \\
$28.038$ & $36.249_{-0.016}^{+0.017}$ & $30.18$ & $34.72$ & $35.78$ & $36.10$ & $36.20$ \\
$30.375$ & $32.896_{-0.016}^{+0.017}$ & $27.86$ & $31.73$ & $32.56$ & $32.79$ & $32.86$ \\
$32.711$ & $30.122_{-0.024}^{+0.016}$ & $25.87$ & $29.21$ & $29.87$ & $30.04$ & $30.09$ \\
$35.048$ & $27.790_{-0.048}^{+0.014}$ & $24.14$ & $27.05$ & $27.59$ & $27.72$ & $27.76$ \\
$37.384$ & $25.801_{-0.080}^{+0.019}$ & $22.63$ & $25.19$ & $25.64$ & $25.74$ & $25.76$ \\ \hline
    \end{tabular}
    \caption{Scattering angle (in degrees) for equal mass, non-spinning BHs with $\Gamma\!=\!1.02264$. $\theta_{n\text{PM}}$ is the analytic PM prediction containing terms up to the at $n$th-order i.e.\ $\theta_{n\text{PM}} = \sum_{i=1}^n \vartheta_{i\text{PM}} \left(\frac{GM}{b}\right)^i$. $\theta_{5\text{PM}}^{1\text{SF}}$ represents $\theta_{5\text{PM}}$ truncated to terms up to order $\nu$ as the $\nu^2$ terms are currently unknown \cite{Driesse:2024xad,Driesse:2024feo}.
    }
\end{table}

\clearpage
\bibliography{biblio}

\end{document}